\documentclass[runningheads,a4paper]{llncs}
\usepackage{amssymb}
\usepackage{graphicx}
\usepackage{listings}
\usepackage{color}
\usepackage{float}
\usepackage{multirow}
\usepackage{lscape}
\usepackage[usenames,dvipsnames]{xcolor}

\usepackage{url}
\usepackage{rotating}
\usepackage{pdflscape}
\urldef{\mailsa}\path|{jean-david.aguilar,safa.jammali, aida.ouangraoua}@USherbrooke.ca|
  
\newcommand{\keywords}[1]{\par\addvspace\baselineskip
\noindent\keywordname\enspace\ignorespaces#1}
\usepackage{lineno} 
\begin{document}

\mainmatter  


\title{Improving spliced alignment for identification of ortholog groups and multiple CDS alignment}

\titlerunning{Spliced alignments for MSA and identification of ortholog groups}

\author{Jean-David Aguilar$^{1,2}$
  \and Safa Jammali$^{1,2}$
  \and Esaie Kuitche$^{1}$
\and A\"ida Ouangraoua$^{1}$}

\authorrunning{Aguilar et al.}

\institute{$^{1}$D\'epartement d'informatique, Facult\'e des sciences, Universit\'e de Sherbrooke, Sherbrooke, Qu\'ebec, Canada\\
  $^{2}$D\'epartement de biochimie, Facult\'e de m\'edecine et des sciences de la sant\'e, Universit\'e de Sherbrooke, Sherbrooke, Qu\'ebec, Canada\\
\mailsa\\
}

%
%

\toctitle{Improving spliced alignment between a priori homologous genes}
\tocauthor{Aguilar et al.}
 \maketitle

\begin{abstract}
The Spliced Alignment Problem (SAP) that consists in finding an optimal semi-global alignment 
of a spliced RNA sequence on an unspliced genomic sequence has been largely considered
for the prediction and the annotation of gene structures in genomes. Here, we re-visit 
it for the purpose of identifying CDS ortholog groups within a set of CDS from
 homologous genes and for computing multiple CDS alignments. We introduce a new 
 constrained version 
 of the spliced alignment problem together with an algorithm that exploits
full information on the exon-intron structure of the input RNA and gene sequences
in order to compute high-coverage accurate alignments. We show how pairwise
spliced alignments between the CDS and the gene sequences of a gene family 
can be directly used in order to clusterize the set of CDS of the gene family 
into a set of ortholog groups. We also introduce an extension
of the spliced alignment problem called Multiple Spliced Alignment Problem (MSAP)
that consists in aligning simultaneously several RNA sequences on several genes
from the same gene family. We develop a heuristic algorithmic solution for the 
problem. We show how to exploit multiple spliced alignments 
for the clustering of homologous CDS into ortholog and close paralog groups,
and for the construction of multiple CDS alignments.
%
An implementation of the method in Python is available on demande to SFA@USherbrooke.ca.

\end{abstract}

\section{Introduction}
\label{intro}

\emph{Spliced alignment} consists in aligning spliced gene products against
unspliced DNA sequences \cite{gelfand1996gene}. It is a key step for genome annotation, 
gene prediction, 
identification of gene structures  and
alternative splicing studies \cite{engstrom2013systematic,zambelli2010}. The quality
of the transcriptome annotation is greatly improved with the use of spliced RNA
data of related genes or genomes. 
In the past two decades, several spliced alignment tools have been developed
for aligning RNA or protein data on genomic DNA (\cite{kapustin2008splign,wu2005gmap} for example).
Such methods look for an
alignment that maximizes the sequence similarity and splice site consensus
signals between a short RNA sequence and a large genome sequence.
They are specifically developed for genome annotation purpose, without
any assumption on the homology of the compared sequences. So, in order to
achieve high sensitivity, they look for high-identity alignments in order
to avoid as much false positive alignments as possible. As a
consequence, these methods perform accurately for the comparison of
sequences from closely related genomes but often misses the alignment
of homologous sequences from distant genomes.

Although spliced alignment methods are classically used for the prediction
and annotation of genes in genomes, they can also be specifically designed
for the purpose of aligning sequences from a priori homologous genes. In
this context, they can be used as a preliminary step for the identification
of orthologous transcripts or proteins isoforms \cite{zambelli2010}, the transfer of transcript annotations between homologous genes 
\cite{blanquart2016assisted} or for studying the
evolution of RNA isoforms in a gene family 
\cite{christinat2013transcript,keren2010alternative,kuitche_2016reconstructing}. Spliced alignment can
also improve the accuracy of multiple homologous CDS alignment by making use
of one or several reference genes.
Current spliced alignment methods make use of the gene
structure by accounting for possible splice sites predicted in the gene
sequence. However, accounting for the exact exon-intron structure of the gene,
when it is known, can help improve the alignment further. Moreover, current
methods do not
make use of the exon structure of the query RNA sequence because they were
mainly designed for aligning EST and RNA-seq data, for which the exon
structure is unknown, on genomic sequences. However, when the RNA sequence
is given with its corresponding gene sequence, it is possible to easily
recover its exon structure and use this information for the spliced alignment
against an other gene.

In this paper, we are interested in the {\sc Spliced Alignment Problem}
 (SAP) that consists in finding an optimal spliced alignment between
a CDS and a gene that captures all sequence
similarities including those between phylogenetically distant homologous
sequences.
For aligning a CDS against a gene, under the assumption of homology between
the sequences, it is possible to make use of the splicing structure of the
sequences in order to accurately detect sequence homologies even in
the case of phylogenetically distant sequences.
First, in Section \ref{spliced}, we introduce a constrained version
of the SAP problem and we present a CDS-gene spliced alignment method that exploits
full information
on the structure of the input CDS and the input gene in order to compute
high-coverage accurate alignments. Second, in Section
\ref{ortholog}, we show how this method can be
directly used for clustering the set of CDS of a gene family into
ortholog and close paralog groups, allowing one-to-one as well as
one-to-many orthology relations. Third, in Section \ref{multiplespliced},  we introduce an extension
of the {\sc Spliced Alignment Problem} called
{\sc Multiple Spliced Alignment Problem  (MSAP)} that consists in finding
an optimal multiple spliced alignment between
a set of CDS and a set of genes. The MSAP problem also extends a homonym
problem introduced in \cite{gelfand1996gene} for the simultaneous spliced alignment
of a set of CDS on a single gene. We describe an algorithm for
the MSAP problem that consists in combining pairwise
spliced alignments into a multiple spliced alignment, by
merging progressively the pairwise alignments while
filtering out false-positive sub-alignments.  
We show that the multiple spliced alignments
allow to identify homologous segments (exons) across a set 
of homologous genes and to compute accurate CDS ortholog groups
and multiple CDS alignments.
Finally, Section \ref{experiment} is devoted to the evaluation of our
method by comparing it to other methods based on the application to the
analysis of set of homologous CDS and genes from the Ensembl-Compara
database. 
\vspace{-.5cm}
\section{Preliminaries: genes, CDS, splicing and orthology}
\label{preliminaries}

In this section, we give some formal definitions that will be useful for the
remaining of the paper. Given a set $S$, $|S|$ denotes the size
of $S$, and given a sequence or an interval $T$, $\texttt{length}(T)$ denotes the
length of $T$.

\noindent {\bf Gene, Exon and CDS:}
A \emph{gene} is DNA sequence on the alphabet of nucleotides
$\Sigma=\{A,C,G,T\}$. Given a gene $G$ of length $n$, an
\emph{exon} of $G$
is a pair of integers $(a,b)$ such that $1\leq a \leq b \leq n$.
The sequence of an exon $(a,b)$ of $G$, denoted by $G[a,b]$, is the
segment of $G$ identified by its start and end positions $a$ and $b$
in $G$.
A CDS $C$ of $G$ is a chain of exons of $G$,
$C = \{(a_1,b_1),\ldots, (a_j,b_j)\}$
 such that for any two
successive exons $(a_i,b_i)$ and $(a_{i+1},b_{i+1})$ in $C$,
$b_i < a_{i+1}$. Thus, the exons of $C$ are non-overlapping and
totally ordered by increasing location on the gene.
We denote by $C[i]$ the i$^{th}$ element of $C$.
We denote the set of introns induced by $C$ by 
$\texttt{Intron}(C) = \{(b_i,a_{i+1}) ~|~ 1\leq i < j\}$.
We denote by $\mathcal{C}(G)$ the set of existing CDS of a gene $G$ and
by $\mathcal{E}(G)$ the set of gene exons of $G$ composing these CDS,
$\mathcal{E}(G)=\bigcup_{C \in \mathcal{C}(G)}{C}$. For example,
if  $|G| = 100$ and $\mathcal{C}(G)$ contains two CDS
$C_1 = \{(11,20), (31,50)\}$ and
$C_2 = \{(11,25), (31,50), (61,90)\}$,
then $\mathcal{E}(G)=\{(11,20), (11,25), (31,50), (61,90)\}$.

The sequence of a CDS $C$ of $G$, denoted
by $G_C$, is the concatenation of the sequences of gene exons composing
$C$ in the order in which they appear in $C$. So, the length of the
CDS sequence $G_C$ equals the sum of the length of the gene exon sequences
composing $G_C$. By extension of the definition of a gene exon, an exon
of a CDS sequence $G_C$ of length $m$ is a pair of integers $(k,l)$ such
that $1\leq k \leq l \leq m$ and the segment $G_C[k,l]$ of $G_C$, identified
by its start and end positions $k$ and $l$ in $G_C$, is exactly one
of the gene exon sequences composing $G_C$.
We denote by $\mathcal{E}(G_C)$ the set of CDS exons composing a CDS sequence
$G_C$. Note that $|\mathcal{E}(G_C)| = |C|$. For the example of
gene $G$ with a CDS $C_1$ given above, $|G_{C_1}| = 30$ and
$\mathcal{E}(G_{C_1}) = \{(1,10), (11,30)\}$.

\noindent {\bf Gene family, orthology relationships:}
  A gene family $\mathcal{G}$ is a set of homologous genes that have derived
  from the same original gene by \emph{duplication} and \emph{speciation}
  events. The genes
  of a gene family are supposed to have similar segments and biochemical
  functions.
  The evolution of a gene family is represented by a rooted binary tree $T$
  whose set of leaves is  $\mathcal{G}$ and internal nodes are ancestral genes that precede an ancestral event labeled as duplications or speciations. A node is an ancestor of a gene
  (leaf)
  if it is on the path between this leaf and the root of the tree.
 The lowest common ancestor (LCA) of two genes is the common ancestor
  to both genes that is the most distant from the root. The orthology
  relationship between genes can be defined from two different
  points of view.
  From a gene tree point of view, two genes of  $\mathcal{G}$ are called
  \emph{orthologs} if their lowest common ancestor in the tree $T$
    is a speciation,
  otherwise they are called \emph{paralogs}. From a similarity point of view,
  orthologous genes have similar sequences, structures and functions. For instance, reciprocal best hits are a common approach for the definition of orthology relations in comparative genomics.

  \noindent {\bf Splicing and spliced alignment:}
  In molecular biology, a eukaryotic mature RNA sequence is obtained
  from a RNA transcript by a phenomenon called \emph{splicing} that consists
  in removing the sequences
  between any two successive exons of a CDS called \emph{introns} and
  joining the resulting exon extremities. 
  The CDS is the segment of the mature RNA sequence that is translated
  into a protein sequence.
  In practice, a full CDS sequence has a length that is multiple of
  $3$ and it starts with a codon \texttt{"ATG"} and ends with a codon
  \texttt{"TAA"}, \texttt{"TAG"} or \texttt{"TGA"}.
  
\begin{figure}[h]
\centering
\includegraphics[height=4.5cm]{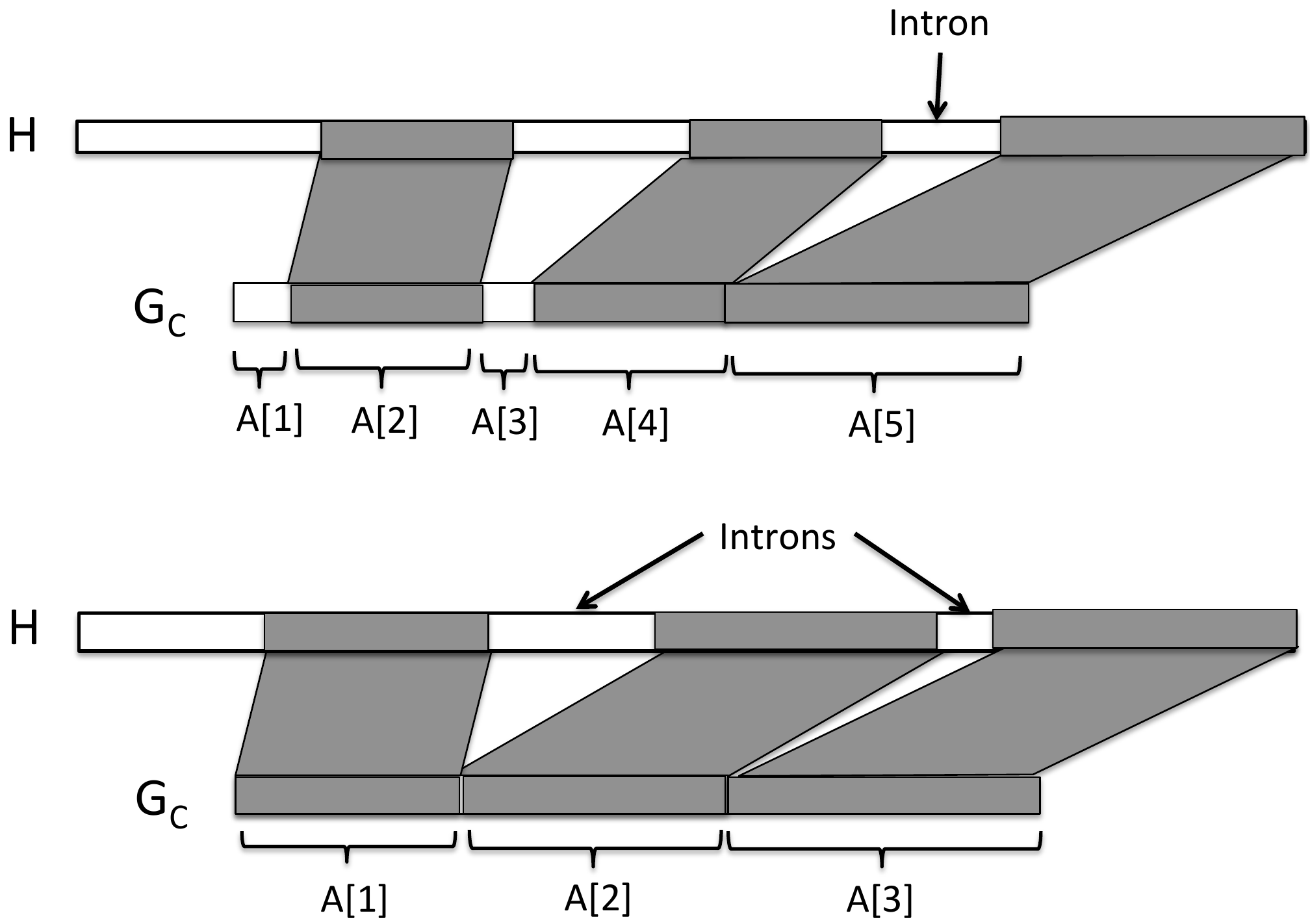}
\caption{{\bf Top.} A spliced alignment between a CDS sequence $G_C$ and
  a gene $H$ composed of $5$ blocks, $3$ conserved blocks ($A[2]$, $A[4]$ and $A[5]$)
  and $2$ deleted blocks ($A[1]$ and $A[3]$), that induce $1$ putative intron.
  {\bf Bottom.} A spliced alignment composed of $3$ conserved blocks
  that induce $2$ putative introns.}
\label{fig:spliced}
\vspace{-.5cm}
\end{figure}

  A \emph{spliced alignment} is an alignment between a CDS and
  a gene sequence that allows to identify conserved exons sequences.
  See Figure \ref{fig:spliced} for example.
Formally, a spliced alignment of a CDS sequence $G_C$ of length $m$
on a gene $H$ of length $n$ is a chain of quadruplets 
$A = \{(k_1,l_1,a_1,b_1),\ldots, (k_j,l_j,a_j,b_j)\}$ called blocks
such that for any block $(k,l,a,b)$ of $A$, $1\leq k \leq l \leq m$,
$a = b = 0$ or $1\leq a \leq b \leq n$, and for any two successive blocks
$(k_i,l_i,a_i,b_i)$ and $(k_{i+1},l_{i+1},a_{i+1},b_{i+1})$,
$l_i = k_{i+1} - 1$.  Moreover, for any two blocks $A[i_1]$ and 
$A[i_2]$ with $i_1 < i_2$, we have $b_{i_1} < a_{i_2}$ if $a_{i_2} \neq 0$. 

The blocks  $(k,l,a,b)$ of $A$ such that $1\leq a \leq b \leq n$
correspond to conserved exon segments $G_C[k,l]$ and $H[a,b]$
between the CDS and the gene sequence. We call them
\emph{conserved blocks} and we denote the set of conserved
blocks of $A$ by
$\texttt{Cons}(A) = \{(k_i,l_i,a_i,b_i) \in A ~|~ b_i \neq 0\}$.
The blocks $(k,l,a,b)$  such that $a = b = 0$
correspond to exon segments $G_C[k,l]$ in the CDS sequence
that are absent in the gene sequence. We call them
\emph{deleted blocks}.
In other terms, the blocks composing $A$ correspond to a chain
of non-overlapping segments of the CDS sequence $G_C$ that are
increasingly located on $G_C$ and cover entirely $G_C$. Moreover
the conserved blocks correspond to a chain of non-overlapping segments
of the gene sequence $H$ that are also increasingly located on $H$.
Thus, there is no crossing in the alignment.
We denote by $A[i]$ the i$^{th}$ block of $A$.

The spliced alignment $A$ also induces a
set of putative gene intron segments.
These intron segments are the gene segments that lie between two
successive blocks of $A$ that are both conserved in the gene sequence.
We denote the set of introns induced by the alignment
$A = \{(k_1,l_1,a_1,b_1),\ldots, (k_j,l_j,a_j,b_j)\}$ by 
$\texttt{Intron}(A) = \{(b_i,a_{i+1}) ~|~ 1\leq i < j ~ \texttt{and} ~ b_i\neq 0 ~\texttt{and} ~ b_{i+1}\neq 0\}$.
Note that if all blocks composing $A$ are conserved, then
$|\texttt{Intron}(A)| = |A| - 1$.

\vspace{-.5cm}

\section{Computation of spliced alignment}
\label{spliced}

In this section, we first re-call two well-known versions of
the spliced alignment problem and we introduce a third version
that we study in this paper. Next, we describe a heuristic algorithm
for the last problem.

\noindent {\bf Spliced alignment problems:}
Given a gene sequence $H$, $H[0,0]$ corresponds to the empty sequence.
Given two nucleotide sequences $S_1$ and $S_2$, let $\texttt{sim}(S_1,S_2)$
denote the score of an optimal global alignment between them. The following
is a reformulation of the less constrained version of the spliced
alignment problem studied in \cite{kapustin2008splign}. This formulation allows a unified
framework to compare the problem with some more constrained versions
of the problem formulated thereafter.\\

\noindent \textsc{Spliced Alignment Problem I  (SAP\_I):}\\
\noindent {\bf Input:} A CDS sequence $G_C$ from a gene $G$ ; a gene sequence $H$. \\
\noindent {\bf Output:} A spliced alignment $A$ of $G_C$ on $H$ that maximizes
$$\sum_{(k,l,a,b)\in A}{\texttt{sim}(G_C[k,l],H[a,b])}$$

The  (SAP\_I) problem only accounts for the alignment scores between the
segments composing the blocks of the spliced alignment. In practice, in more
than 99$\%$ of real cases of splicing,
the removed intron sequences start with a dinucleotide sequence
\texttt{"GT"} and ends with a dinucleotide sequence \texttt{"AG"} respectively
called canonical donor and acceptor splice sites \cite{burset2000analysis,sparks2005incorporation}. 
Thus, in order to
improve the accuracy of the spliced alignment, a more constrained
version of the problem allows to account for the intron segments induced
by an alignment. Given a nucleotide
sequence $S$, let $\texttt{intr}(S)$ denote the intron score of $S$
accounting for the presence or absence of canonical splice sites at the
extremities of $S$. A sequence with two canonical splice sites at its
extremity has a higher score than a sequence with only one  which
has a higher score than a sequence without canonical splice sites.
A more constrained version of the spliced
alignment problem studied in \cite{kapustin2008splign} is the following.\\

\noindent \textsc{Spliced Alignment Problem II  (SAP\_II):}\\
\noindent {\bf Input:} A CDS sequence $G_C$ from a gene $G$ ; a gene sequence $H$. \\
\noindent {\bf Output:} A spliced alignment $A$ of $G_C$ on $H$ that maximizes
$$\sum_{(k,l,a,b)\in A}{\texttt{sim}(G_C[k,l],H[a,b])} + \sum_{(b,a)\in \texttt{Intron}(A)}{\texttt{intr}(H[b,a])}$$

The SAP\_I and SAP\_II problems do not account for the exon structure of the CDS sequence and the exon-intron structure of the gene
sequence. In order to further improve the accuracy of the spliced
alignment, we consider a more constrained version of the problem that
accounts for the actual exons in the CDS and the gene sequence.
Given a conserved block $(k,l,a,b)$ of a spliced alignment,
let $\texttt{exon}_{\mathcal{E}(G_C),\mathcal{E}(H)}(k,l,a,b)$
denote the score of a conserved block accounting for the correspondence
of the block with an actual exon in the CDS or in the gene sequence.
A block that corresponds to an exon in both sequences has a higher score
than a block with an exon correspondence in
only one of the sequences which has a higher score than a block without
any exon correspondence.
The more constrained version of the problem is defined as follows.\\

\noindent \textsc{Spliced alignment Problem III (SAP\_III):}\\
\noindent {\bf Input:} A CDS sequence $G_C$ from a gene $G$ ;  the set of exons
$\mathcal{E}(G_C)$ of $G_C$ ; a gene sequence $H$ ; the set of exons
$\mathcal{E}(H)$ of $H$. \\
\noindent {\bf Output:} A spliced alignment $A$ of $G_C$ on $H$
that maximizes
$$\sum_{(k,l,a,b)\in A}{\texttt{sim}(G_C[k,l],H[a,b])} + \sum_{(k,l,a,b)\in \texttt{Cons}(A)}{\texttt{exon}_{\mathcal{E}(G_C),\mathcal{E}(H)}(k,l,a,b)}$$
  $$+ \sum_{(b,a)\in \texttt{Intron}(A)}{\texttt{intr}(H[b,a])}$$

\noindent {\bf Heuristic algorithm for the (SAP\_III) problem:}
The general approach followed by most heuristic spliced alignment methods
for the SAP\_I and SAP\_II problems consists in first computing high-identity
local alignments between the CDS and the gene sequence. In a second step,
these local alignments are used as anchors and a global alignment algorithm
is applied in the regions between the anchor alignments in order to complete
the spliced alignment.
%
Our heuristic spliced alignment method for the SAP\_III problem also starts
with the computation of highly conserved local alignments used as anchors.
However, in a second step, we use the exon structure of the CDS sequence
to extend the anchors towards the extremities of the
CDS exons containing them. This steps drastically reduces the length of the CDS
sequences remaining between the extended anchor in the deleted blocks.
In a third step, we apply a global alignment in the remaining regions.
In a fourth step,
we correct the splicing junctions in order to recover GT$\slash$AG canonical
splice sites that were missed because of the insertion or deletion of
nucleotides in the exon sequences, resulting in gaps in the block alignments.
Finally, in a fifth step, we use the exon-intron structure
of the gene sequence to further correct the block extremities. The details of
the steps composing the method are given below.

\begin{figure}[h]
\centering
\includegraphics[height=5.5cm]{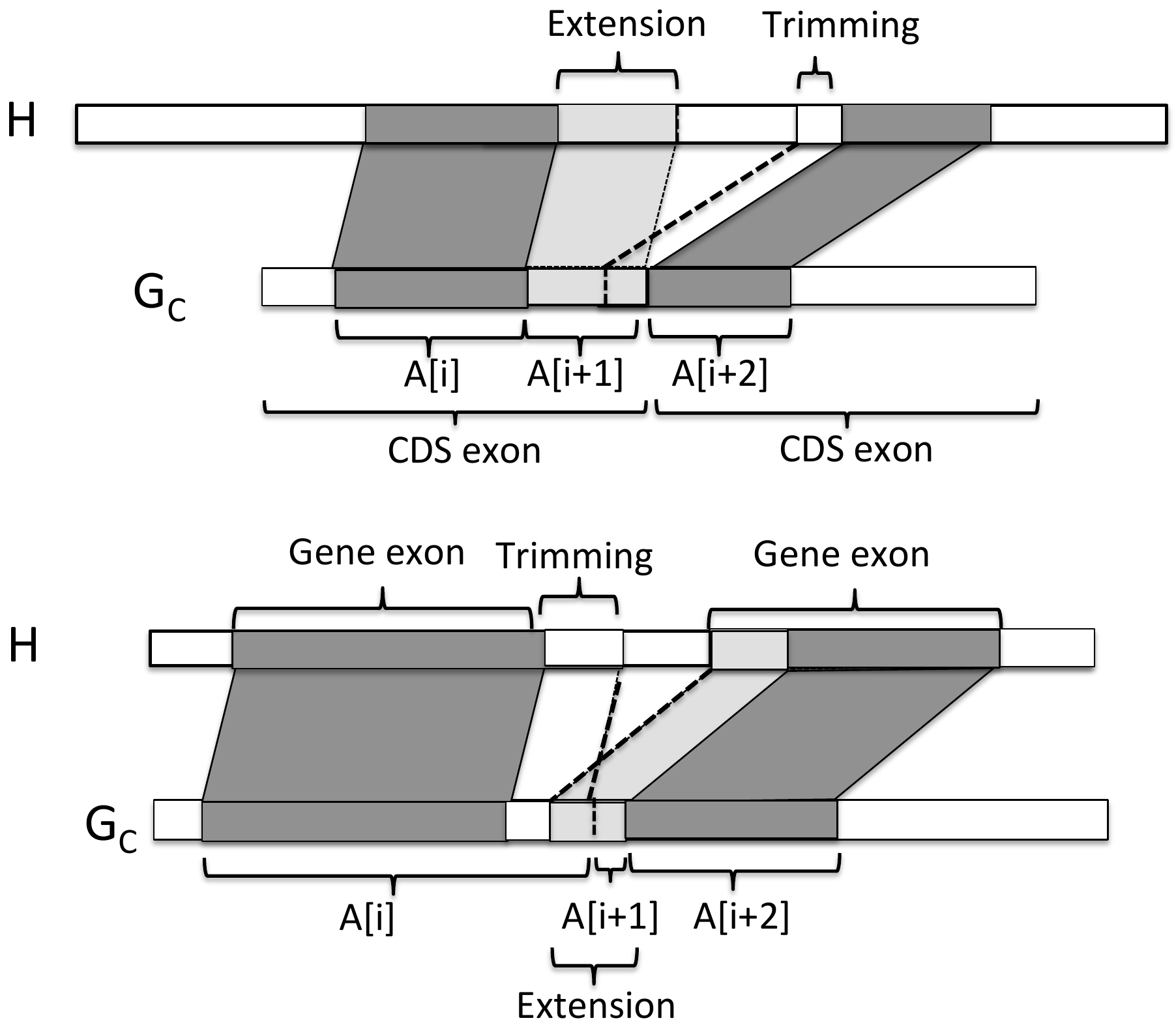}
\vspace{-.3cm}
\caption{{\bf Top.} Illustration of Step 2 of the heuristic spliced alignment
  method: the right extremity of the block $A[i]$ is extended according to the
  first CDS exon, and that requires the trimming of the left extremity of
  the block $A[i+2]$.
  {\bf Bottom.} Illustration of Step 5: the right extremity of the block $A[i]$
  is trimmed according to the
  first gene exon, and next the left extremity of the block $A[i+2]$
  is extended according to the second gene exon.}
\label{fig:computespliced}
\vspace{-.5cm}
\end{figure}

\begin{description}
\item {\bf Step 1. Local alignment.} This step is achieved using Translated
  Blast (tblastx) 
    with an E-value threshold of $10^{-2}$ in order to obtain local alignments.
    Tblastx is used in order to account for the translation of the sequences
    into amino acid sequences. This allows to detect amino acid sequence
    conservation even in the presence of translational frameshifts.
    
   \item {\bf Step 2. Block extension based on CDS exons.} This step is
     repeated iteratively in order to prioritize highly conserved local
     alignments. Local alignments are classified into four groups according
     to their E-value scores, less than $10^{-7}$, between $10^{-7}$ and
     $10^{-5}$, between $10^{-5}$ and $10^{-3}$ and greater than  $10^{-3}$.
     For each group from the more to the less conservative, a maximum size
     set of pairwise compatible anchors is kept as conserved blocks
     and the blocks are extended as follows. Given a block
     extremity that does not correspond to a CDS exon extremity, the block
     extremity is extended until the closest
     CDS exon extremity if this extension does not decrease the percentage of
     nucleotide identity in the block alignment. If the extension induces
     an overlapping with another conserved block on the CDS sequence, the
   other block is trimmed in order to free space for the extension.
     In this case, the extension and the trimming are applied
     only if they strictly increases the nucleotide identity of the two
     block alignments. See Figure \ref{fig:computespliced} for an illustration.
     
   \item {\bf Step 3. Global alignment of remaining regions.} This step
     consists in applying a classical global alignment algorithm for
     the regions of the CDS in deleted blocks, not yet covered by the
     spliced alignment. The conserved blocks induced by the global
     alignment are added in the spliced alignment.
     For each comparison, we make use of an exact
     semi-global alignment
     that penalizes no end gaps in the two compared sequences.
     
   \item {\bf Step 4. Correction of exon junctions.} In this step, 
     the  introns induced by the spliced alignment are refined.
     Given an induced intron $(b,a)$ between two conserved blocks
     of the alignment, if the donor and acceptor
     splice sites of the intron are not both canonical, we look for a shift
     of the block junction that corresponds to a pair of canonical splice
     sites and does not decrease the percentage of nucleotide identity
     in the flanking block alignments. The space search is limited
     to a maximum of 30-nucleotide shift and 3-codon gaps.
     
   \item {\bf Step 5. Correction of block extremities based on gene exons.}
     This step is composed of two phases. First, each extremity of a conserved
     block that does not correspond to a gene exon extremity
     or a CDS exon extremity is trimmed until the closest gene exon extremity
     if there exists such a gene exon extremity. This allows to free space
     in order to extend neighboring conserved block extremities in a second
     phase.
     In the second phase, each conserved block extremity that still does
     not correspond
     to a gene exon extremity or a CDS exon extremity is extended
     until the closest available gene exon extremity if this extension
     does not decrease the nucleotide identity of the block alignment.
     See Figure \ref{fig:computespliced} for an illustration.
\end{description}

Note that, the algorithm used for each step of the method
can be replaced by any other algorithm solving the same problem.
For instance, Step 1 can use a faster local alignment algorithm than
tblastx. Steps 3 and 4 can be merged by using a global alignment
algorithm that accounts simultaneously for the sequence similarities and
the intron scores such as the global alignment algorithm developed
\cite{kapustin2008splign}.
The output of the spliced alignment method is a spliced alignment whose
blocks extremities maximize the correspondence with the exon extremities
of the input CDS and gene sequences.

\section{Identification of CDS ortholog groups}
\label{ortholog}

\vspace{-.2cm}
In this section, we give a definition of CDS ortholog groups based on pairwise
spliced alignments. We first start with a definition of orthologous CDS.

In \cite{zambelli2010}, an extension of the concept of 
gene orthology to spliced transcript orthology
  was introduced. They defined orthologous transcripts as two
  structurally similar transcripts from two orthologous genes. In \cite{kuitche_2016reconstructing},
  we have introduced a new protein tree model of transcript evolution along gene
  trees, that is similar to the gene tree model of gene evolution along
  species tree with duplication and speciation events. In the model of
  transcript evolution along gene trees, we have introduced a third type of
  event called \emph{creation} representing the separation of
  two lineages of structurally different transcripts.
  This model have led to an extended definition of protein and transcript
  orthology, based
  on the protein tree model and relaxing the constraint that two orthologous
  transcripts should come from orthologous genes. We call two transcripts
  ortholog if their LCA in the protein tree is a speciation or a duplication
  (not a creation), otherwise they are called paralogs. From a similarity
  point of view, the corresponding definition of transcript orthology
  is that two transcripts
  are orthologs if they are structurally similar and come from two homologous genes, not necessarily orthologous genes. Next, two orthologous transcripts are
  \emph{ortho-orthologs} if they come from orthologous genes, and
  \emph{para-orthologs} otherwise.
  Note, that our definition of ortho-orthologs then corresponds
  to the definition of orthologous transcripts from \cite{zambelli2010}.
  Thus, the orthology relationship between two transcripts from two homologous
  genes relies on the evaluation of the structural similarity between the
  transcripts. Here, we evaluate this structural similarity using the CDS
  associated to the transcripts.\\
%

  \noindent {\bf CDS orthology:}
  Let $C_1$ and $C_2$ be two CDS from two homologous genes $G$ and $H$
  respectively. Let $A_1$ be a spliced alignment
  of the CDS sequence $G_{C_1}$ on the gene sequence $H$, and $A_2$
  a spliced alignment of $H_{C_2}$ on $G$. Using $A_1$ and $A_2$, we define $C_1$ and $C_2$ as orthologs if:\\
  (1) $|~C_1~| = |~C_2~|$ and \\
  (2) $\texttt{Intron}(A_1) = \texttt{Intron}(C_2)$
  or $\texttt{Intron}(A_2) = \texttt{Intron}(C_1)$ and \\
  (3) for any $i, 1\leq i \leq ~|~C_1~|$,
  $[\texttt{length}(C_1[i]) - \texttt{length}(C_2[i])]~ \% ~3 = 0$.

  In other terms, $C_1$ are $C_2$ as orthologs if (1) they have the same
  number of exons, (2) the spliced alignment of $C_1$ on $H$ induced
  the same introns as $C_2$ in $H$ or
  the spliced alignment of $C_2$ on $G$ induces the same introns
  as $C_1$ in $G$ and (3) the lengths of each pair of corresponding exons
  in $C_1$ and $C_2$ should be congruent modulo $3$. The conditions
  (1) and (2) ensure that the two CDS have the same exon structure.
  The condition (3) ensures that the two CDS are translated in the
  same codon phase in each pair of corresponding exons.
   
  Note that this definition only requires that one of the spliced
  alignments $A_1$ and $A_2$ supports the orthology relation.
  An alternative more stringent definition of CDS orthology
  consists in requiring the reciprocity, i.e. that
  $A_1$ and $A_2$ both support the orthology relation,
  by using the 'and' statement instead of the 'or' statement
  in the condition (2).\\

  \noindent {\bf CDS ortholog groups:} Given a set of CDS
  $\mathcal{C}$ from a set of homologous genes $\mathcal{G}$,
  the transitivity of the CDS orthology relation is used
  to identify distant orthologs and co-orthologs in $\mathcal{C}$
  that cannot be identified by means of the CDS structural
  similarity. Such orthologs are typically missed because
  of partial spliced alignments due to low sequence similarity.
  
  The CDS orthology relation on $\mathcal{C}$ is then extended
  into an equivalence relation such that for any three CDS
  $C_1,C_2,C_3$ in $\mathcal{C}$, if $C_1$ and $C_2$ are
  orthologs and $C_2$ and $C_3$ are orthologs, then $C_1$ and
  $C_3$ are also orthologs.
  The CDS ortholog groups are defined as the equivalence
  classes of the resulting equivalence relation.

 \section{Computation of multiple spliced alignment}
 \label{multiplespliced}

 The concept of spliced alignment was extended 
 in order to define multiple spliced alignments \cite{brendel2004gene,gelfand1996gene}.
 A multiple spliced alignment was then defined as a 
 simultaneous spliced
 alignment of several RNA sequences on a single
 gene sequence.  Multiple spliced alignment allows to improve
 the accuracy of gene structure prediction by making use of
 several RNA targets simultaneously. 
 Here, we further extend the concept in order to consider the simultaneous
 spliced alignment of several CDS sequences on several genes. See Figure
 \ref{fig:multiplespliced} in Appendix for example.
 
 Based on the formalism introduced in Section \ref{preliminaries}, we defined
 a \emph{multiple spliced alignment} of a set of CDS sequences $\mathcal{C}$
 on a
 set of genes $\mathcal{G}$ as a chain $A = \{A[1], \ldots, A[j]\}$ of sets
 called  multi-blocks such that $A[i]$ denotes the i$^{th}$ multi-block of
 $A$. Each multi-block $A[i]$ is a set of pairs
 $A[i] = \{(s_i^{x},e_i^{x})  ~|~ x \in \mathcal{C}\cup\mathcal{G}\}$ such
 that for any sequence $x\in \mathcal{C}\cup\mathcal{G}$,
 $s_i^{x} = e_i^{x} = 0$ or $1\leq s_i^{x}  \leq e_i^{x} \leq \texttt{length}(x)$.
 For any two multi-blocks $A[i_1]$ and $A[i_2]$ with $i_1 < i_2$, and
 any sequence $x\in \mathcal{C}\cup\mathcal{G}$, $e_{i_1}^{x} < s_{i_2}^{x}$ if
 $s_{i_2}^{x} \neq 0$. Moreover, for any CDS sequence $x\in \mathcal{C}$,
 the set of segments of $x$ induced by $A$,
 $\{x[s_i^{x},e_i^{x}]  ~|~  1\leq i \leq j ~\texttt{and}~ e_i^{x} \neq 0\}$
 cover  entirely $x$.

 A multi-block
 $A[i] = \{(s_i^{x},e_i^{x})  ~|~ x \in \mathcal{C}\cup\mathcal{G}\}$
 corresponds to a group of segments (one segment $x[s_i^{x},e_i^{x}]$
 for each sequence $x$ such that $e_i^{x} \neq 0$) that are
 homologs and absent in any sequence $x$ such that 
 $e_i^{x} = 0$. By definition,
 the  multi-blocks composing $A$ are non-overlapping in any sequence
 and increasingly located on each of the sequence. Thus, there is
 no crossing in the alignment.

 A multiple spliced alignment $A = \{A[1], \ldots, A[j]\}$ of a set of CDS
 sequences $\mathcal{C}$  on a set of genes $\mathcal{G}$ induces a spliced
 alignment $A_{x,y}$ for each pair $(x,y) \in \mathcal{C}\times \mathcal{G}$.
 The induced spliced alignment consists of a reduction of the multi-blocks
 $A[i]$, $1\leq i \leq j$  such that $e_i^{x} \neq 0$, i.e the multi-blocks
 in which the segment of the CDS sequence $x$ is not empty,
 $A_{x,y}= \{(s_i^{x},e_i^{x},s_i^{y},e_i^{y}) ~|~ 1\leq i \leq j ~\texttt{and}~ e_i^{x} \neq 0\}$.

 Given a spliced alignment $A_{x,y}$ of a CDS sequence $x$ on a gene sequence
 $y$, let $\mathbb{S}(A_{x,y})$ denote the score of the spliced alignment
 $A_{x,y}$ as defined for the SAP\_I, SAP\_II or
 SAP\_III  problem. We defined the corresponding multiple spliced alignment
 problem as follows:\\

\noindent \textsc{Multiple spliced alignment Problem (MSAP\_$\mathbb{S}$):}\\
\noindent {\bf Input:} a set of CDS sequence $\mathcal{C}$ ; a set of genes $\mathcal{G}$. \\
\noindent {\bf Output:} A multiple spliced alignment $A$ of
$\mathcal{C}$ on $\mathcal{G}$ that maximizes the sum of the scores of
induced pairwise spliced alignments:
$\sum_{(x,y)\in \mathcal{C}\times \mathcal{G}}{\mathbb{S}(A_{x,y})}$.\\

\noindent {\bf Heuristic algorithm for the MSAP\_III problem:}
We now describe a heuristic algorithm for building a multiple spliced
alignment of a set of CDS sequence $\mathcal{C}$  on a set of
genes $\mathcal{G}$, given the spliced alignments for each
pair of sequences  $(x,y)\in \mathcal{C}\times \mathcal{G}$
computed using a pairwise spliced alignment method.

The idea behind the algorithm is to progressively merge the blocks
of the input pairwise spliced alignments into multi-blocks of
the target multiple spliced alignment. In order to merge a
block into an existing multi-block, we rely on the compatibility
between the segments composing the blocks. Given a sequence $x$,
and two segments of $x$, $x[s_1,e_1]$ and $x[s_2,e_2]$, we say that
the segments are  \emph{compatible} if $s_1 = s_2 \pm \epsilon$ and
$e_1 = e_2$ or $s_1 = s_2 $ and $e_1 = e_2\pm \epsilon$.
In other terms, the segments are compatible if they are nested
with an equality of at least one of the two extremities of the segments,
and the other extremities have a difference of at most $\epsilon$.
If $x[s_1,e_1]$ and $x[s_2,e_2]$ are compatible, we denote
by $\max((s_1,e_1),(s_2,e_2))$ the largest of the two intervals.
In practice $\epsilon$ is set to $50$. 

Next, we define the compatibility between a pairwise block and
multi-block, and the compatibility between two multi-blocks as follows.
Let $A$ be a multiple spliced alignment of $\mathcal{C}$ on
$\mathcal{G}$ obtained at some step of the algorithm.
Let $(k,l,a,b)$ be a block in a spliced alignment of a pair
of sequence $(x,y)\in \mathcal{C}\times \mathcal{G}$. The block
$(k,l,a,b)$ is \emph{compatible} with a multi-block
$A[i] = \{(s_i^{x},e_i^{x})  ~|~ x \in \mathcal{C}\cup\mathcal{G}\}$ of
$A$  if (1) the CDS segments  $x[k,l]$ and
$x[s_i^{x},e_i^{x}]$ are compatible or (2) the gene segments
$y[a,b]$ and $y[s_i^{y},e_i^{y}]$ are compatible.
Thus, the  block  $(k,l,a,b)$  and the
multi-block $A[i]$ are compatible if their corresponding segments
in the CDS sequence $x$ or in the gene sequence $y$ are compatible.
Similarly, two multi-blocks $A[i]$ and $A[j]$ of $A$ are \emph{compatible}
if for any sequence $x\in \mathcal{C}\cup \mathcal{G}$,
the segments $x[s_i^{x},e_i^{x}]$ and $x[s_j^{x},e_j^{x}]$
are compatible.

The algorithm starts with an empty multiple spliced alignment $A$,
and considers all conserved blocks contained in the pairwise spliced
alignments iteratively. Let $(k,l,a,b)$ be a
block in the spliced alignment of $x\in \mathcal{C}$ on $y \in \mathcal{G}$
considered at some iteration of the algorithm.
By construction, the block $(k,l,a,b)$ can be compatible with zero,
one or two multi-blocks of $A$. Depending on the case, the multiple
spliced alignment $A$ is refined as follows. 

\begin{description}
\item {\bf Case 1.} If the block $(k,l,a,b)$ is not compatible with
  any multi-block, then a new multi-block $A[i]$ is added such that
  $(s_i^{x},e_i^{x}) = (k,l)$, $(s_i^{y},e_i^{y}) = (a,b)$ and for any
  other sequence $z \in \mathcal{C}\cup \mathcal{G} \setminus\{x,y\}$,
  $(s_i^{z},e_i^{z}) = (0,0)$.

  \item {\bf Case 2.} If $(k,l,a,b)$ is compatible with
    a single multi-block $A[i]$, then the block $(k,l,a,b)$ is added to
    the multi-block $A[i]$ such that
    $(s_i^{x},e_i^{x}) = \max((k,l), (s_i^{x},e_i^{x}))$  and
    $(s_i^{y},e_i^{y}) = \max((a,b), (s_i^{y},e_i^{y}))$.

  \item {\bf Case 3.} If $(k,l,a,b)$ 
is compatible with two multi-blocks  $A[i]$ and $A[j]$,
one $A[i]$ satisfying the compatibility condition (1) for the CDS sequence $x$
and the other one $A[j]$ satisfying the compatibility condition (2) for the gene sequence
$y$, there are two possibilities.

\begin{description}
\item {\bf Case 3.a.}  If the multi-blocks $A[i]$ and $A[j]$
  are compatible, then they are merged into a single multi-block $A[u]$
  replacing $A[i]$ and $A[j]$, such that for each sequence
  $z\in \mathcal{C}\cup \mathcal{G}$,  the segment of $z$ included
  in the new multi-block is
  $(s_u^{z},e_u^{z}) = \max((s_i^{z},e_i^{z}), (s_j^{z},e_j^{z}))$.

\item {\bf Case 3.b.}  If the multi-blocks  $A[i]$ and $A[j]$ are not
  compatible, then there are 3 cases: either the block $(k,l,a,b)$
  is erroneous and it should be discarded, or the occurrence of the
  segment $(s_i^{x},e_i^{x})$ in $A[i]$ is erroneous and it should
  be discarded, or the occurrence of the segment $(s_i^{y},e_i^{y})$
  in $A[j]$ is erroneous and it should be discarded. One should be
  very careful in choosing one of these cases in order to avoid
  subsequent conflicts caused by a wrong decision at this step.
  In order to decide of the correctness of the block, the percentage
  of nucleotide identity in the block alignment $(k,l,a,b)$ is one
  criteria but it is not sufficient. Indeed, when an exon is
  duplicated in a gene (as we will see in one example in
  Section \ref{experiment}), a block with a high percentage
  of nucleotide identity in the alignment can be erroneous.
  Consequently, we give a very conservative definition of
  a correct block in the case of conflict. We say that the block
  is \emph{correct} if the percentage of nucleotide identity in its
  alignment is greater or equal to a threshold $\tau$ and the
  block alignment contains no gaps. In practice $\tau$ is set to $60\%$.

\begin{description}
 \item {\bf Case 3.b.i.}  If the block $(k,l,a,b)$  is not
   correct then it is not added to the multiple spliced alignment.
 \item {\bf Case 3.b.ii.} Otherwise, we define some scores for
   the occurrence of the  segment $(s_i^{x},e_i^{x})$ in $A[i]$
   and the occurrence of the segment $(s_i^{y},e_i^{y})$ in $A[j]$.
   The occurrence whose score is the higher is kept and the block
   $(k,l,a,b)$ is added in the corresponding multi-block, while
   the other occurrence is discarded from its multi-block.

   The score of the occurrence of the segment $(s_i^{x},e_i^{x})$
   in $A[i]$ equals the number of pairwise spliced alignment blocks
   supporting the occurrence of $(s_i^{x},e_i^{x})$ in $A[i]$.
   Let $y$ be any gene sequence that has an occurrence
   in $A[i]$, a block in the pairwise spliced alignment of $x$
   on $y$ is a support if it is compatible with the multi-block
   $A[i]$. The score of the occurrence of the segment $(s_i^{y},e_i^{y})$
   in $A[j]$ is defined similarly based on the blocks of
   the pairwise spliced alignment of any CDS sequence $x$
   present in $A[j]$ on the gene sequence $y$.
\end{description}
  
\end{description}
\end{description}
  
  The output of this algorithm is a multiple spliced alignment whose
multi-blocks extremities maximize the correspondence with the exon extremities
of the input CDS and gene sequences. We now describe how to use a multiple spliced alignment in order
to cluster the CDS of set of homologous genes into groups
of orthologs and close paralogs.

  \noindent {\bf CDS ortholog and close paralog groups:} 
  The multiple spliced alignment $A =  \{A[1], \ldots, A[j]\}$
  of a set of CDS sequences
  $\mathcal{C}$ on a set of genes $\mathcal{G}$ is 
  used to define clusters of CDS orthologs, co-orthologs 
  and close paralogs as follows. Let $x$ and $y$ be two CDS sequences   in $\mathcal{C}$ from two genes $G$ and $H$ in $\mathcal{G}$, possibly the same gene $G=H$. The CDS sequences $x$ and $y$ 
   belong to the same group if for any multi-block 
   $A[i] \in A$, (1) either $(s_i^{x},e_i^{x}) = (s_i^{y},e_i^{y}) = (0,0)$  or $e_i^{x} \neq 0$ and  $e_i^{y} \neq 0$ and 
   (2)  $(e_i^{x} - s_i^{x}) - (e_i^{y} - s_i^{y})~ \% ~3 = 0$.
   If $G=H$, $x$ and $y$ are called \emph{close paralogs}.
   Otherwise, $x$ and $y$ are \emph{co-orthologs} or 
   \emph{orthologs}.
  
  \noindent {\bf Multiple CDS alignment:}
  The multiple spliced alignment $A$ of the set of CDS sequences
  $\mathcal{C}$ on the set of genes $\mathcal{G}$ is also
  used to define a Multiple Sequence Alignment (MSA) 
  of the CDS sequences and the genes exon segments. 
  For each multi-block 
   $A[i] \in A$, the set of segments 
   $\{x[s_i^{x},e_i^{x}] ~ | x \in \mathcal{C}\cup \mathcal{G}~\}$
   is aligned using a MSA tools, and the resulting 
   alignments
  are concatenated in order to obtain a global multiple     
  alignment of all the CDS sequences with the concatenation 
  of the gene exon segments. In practice, we use the sequence aligner Muscle \cite{edgar2004muscle} for the multiple alignment of the segments composing a multi-block.

  \vspace{-.5cm}
\section{Application}
\label{experiment}

 We applied our algorithms on two sets of homologous genes from the Ensembl-Compara database release 89 \cite{cunningham2015ensembl}. We 
 wanted to evaluate the aptitude of the methods (1) to compute
 pairwise high CDS-coverage spliced alignments maximizing 
 the correspondence with  
 actual exon extremities in the CDS and gene sequences, (2)
 to identify accurate CDS ortholog groups covering phylogenetically
 distant genes based on pairwise spliced  alignments, and 
 (3) to compute accurate multiple CDS alignments, and ortholog and close paralog groups based on multiple spliced  
 alignments. In the remaining of the section, our methods
 are named SpliceFAmAlign (SFA).\\

\noindent {\bf Dataset:}
The dataset contains $16$ genes with their CDS sequences from two gene families, FAM86  and MAG: $8$ genes per family, $14$ CDS for FAM86 and $26$ for MAG. For each family, the genes are from seven different amniote species which are \emph{human, chimpanzee, mouse, rat, cow and chicken} (for FAM86) or \emph{lizard} (for MAG). In particular, the $3$ MAG human genes contain a pair of duplicated
exons  separated by a third smaller exon.
Table \ref{familiesdetails} in Appendix gives more details about the dataset.\\

\noindent {\bf 1. Evaluation of the pairwise spliced alignments:}
We compared the results of our pairwise spliced alignment algorithm
SFA for the {\sc SPA\_III} problem with the results of the current best tool SPlign \cite{kapustin2008splign} developed for the {\sc SPA\_II} problem. We wanted to compare the ability of the methods to correctly identify actual 
exon extremities in the CDS and the gene sequence, canonical splice
sites and actual introns in the gene sequence. We also evaluated 
the CDS coverage of the spliced alignments computed using the two methods. For the two methods, we considered different values for a  minimal block nucleotide identity parameter called \texttt{min\_idty} 
such that the blocks with lower identity ratio than this threshold 
in their alignment 
are discarded from the spliced alignment. For SPlign, we tested
the values $0.0$ and $0.75$ (default parameter), and for SFA
the values $0.0$, $0.6$ (default parameter), $0.7$, and $0.75$ for 
\texttt{min\_idty}. \\
%
The full results are shown in Appendix in Table \ref{tab:structure}. As expected, the spliced alignments computed 
by SFA cover a larger percentage of the CDS sequences. For the 
ability to correctly recover actual exon extremities in the CDS and 
the gene sequences, SFA recovers almost twice more actual exon 
extremities than SPlign, but its number of blocks is also doubled. 
Thus, the ratios of actual CDS or gene exons extremities, and the
ratio of canonical splice sites is lower for SFA.\\

\noindent {\bf 2. Evaluation of the CDS ortholog groups computed based on pairwise spliced  alignments:}
We applied our ortholog clustering method using  pairwise spliced 
alignment in order to obtain
a set of ortholog groups based on structural similarity. We also computed a set of ortholog clusters using the OrthoMCL clustering tool \cite{li2003orthomcl} based on sequence similarity. We wanted
to evaluate the ability of the SFA  method to identify accurate CDS ortholog groups covering phylogenetically distant genes, and to compare the structure-based ortholog clusters obtained
with the SFA method to the clusters obtained using a
sequence-based clustering method. 

The sets of ortholog clusters computed by the two methods are depicted in 
Figures \ref{fig:fam86_clusters} and \ref{fig:mag_clusters} in Appendix. 
For FAM86, SFA allows to recover a $5$-CDS ortholog group composed of CDS conserved across 
2 \emph{human}, 1 \emph{mouse}, 1 \emph{rat} and 1 \emph{cow} genes. It also returns a pair of orthologous CDS from
2 orthologous \emph{human} and \emph{chimpanzee} genes.
 For the MAG family, SFA allows 
to recover a $3$-CDS ortholog group with CDS from
1 \emph{human}, 1 \emph{rat} and 1 \emph{cow} genes,
and a pair of orthologous CDS from
2 orthologous \emph{human} and \emph{rat} genes.
The multiple alignment of the corresponding protein
sequences confirms the accuracy of the ortholog groups 
(data not shown). The structural similarity between
the orthologous CDS is also highlighted in Figures
\ref{fig:fam86_clusters} and \ref{fig:mag_clusters} (Appendix).
Concerning the comparison between the OrthoMCL and SFA
clusters, for FAM86, SFA returns $9$ clusters 
and OrthoMCL returns $5$ clusters. For MAG, 
SFA returns $23$ clusters while OrthoMCL returns $3$ clusters. 
In both cases, the two sets of clusters are compatible as
the clusters computed by the SFA methods are included in larger clusters 
computed by OrthoMCL.  Figures
\ref{fig:fam86_clusters} and \ref{fig:mag_clusters} (Appendix) 
reveals
that the SFA clusters merged by OrthoMCL share some sequence 
similarities indeed, but they also have different exon structures.\\


\noindent {\bf  3. Evaluation of the multiple spliced alignment:}
We applied our multiple spliced alignment algorithm 
using the pairwise spliced alignments as input. Next, we 
computed a set of ortholog and close paralog groups,
and a multiple alignment of the CDS sequences using
the methods described in Section \ref{multiplespliced}.
First, we wanted to evaluate the accuracy of the clusters
reconstructed based on the multiple spliced alignment
and their compatibility with the clusters
reconstructed using the more stringent method
based on pairwise spliced alignments. For both families,
all pairwise SFA clusters are included in larger
clusters defined using the multiple SFA clustering method
(data not shown).
Second, we wanted to evaluate the accuracy of the
reconstructed multiple CDS alignments. To do so, we 
compare them with the multiple CDS alignments obtained using
the only existing tools MACSE \cite{ranwez2011} 
that allows to align 
multiple CDS sequences while accounting for translational 
frameshifts. MACSE \cite{ranwez2011} was ran with its default parameter values to compute a multiple CDS alignment for each of the gene families.
For each gene family, we compare
the two CDS multiple alignments, the one from our method
SFA and the one from MACSE, based on two criteria which are the percentage of column identity in the alignment and the ratio of long gaps in the alignment that correspond to real exon-exon junctions in the CDS sequences. We define a \emph{long gap}
as a gap of length greater or equal to $20$ between  two nucleotides 
$x[i]$ and $x[i+1]$ of a CDS sequence $x$
in the multiple CDS alignment. This gap corresponds
to a real exon-exon junction in the CDS if $x[i]$ and $x[i+1]$ belong
to two different (consecutive) exons in $x$.

The results are shown in Table \ref{msaevaluation} in Appendix. In all cases, SFA recovers
a higher number and a higher ratio of long gaps corresponding to real exon junctions in the CDS sequences.

\vspace{-.5cm}
\section{Conclusion}
\label{conclusion}

This paper introduces a new constrained version and an extended version
of the Spliced Alignment Problem for the purpose of identifying CDS
ortholog groups within a set of CDS from homologous genes, and
for computing multiple CDS alignments. We have developed a variety
of algorithmic solutions for the computation of pairwise and
multiple spliced alignments. We show how this framework can
be used to improve the definition of CDS orthology clusters and
multiple CDS alignments. The application of the algorithms
to real datasets shows that the framework is particularly
useful for identifying CDS ortholog groups that
are conserved accross phylogenetically distant genes.

On the algorithmic side, the new problems presented require
a more in-depth investigation. For instance, the heuristic
algorithm described for the constrained spliced alignment problem
could make use of a exact global alignment method
with scoring schemes defined to account for all the
constaints of the alignments. The heuristic algorithm
described for the multiple spliced alignment problem follows
a classical greedy approach used by most multiple sequence
aligner based on pairwise alignment. What is the complexity
of the problem ? It can be seen as a problem of reconstructing
a macroscopic multiple alignment and then it is likely to
be NP-hard. If so, it is possible that the problem is fixed-parameter
tractable with respect to a parameter such as the maximum
number of blocks in a pairwise alignment or the maximum size
of a conflicting set of blocks. Future work will also make
use of benchmark datasets in order to confirm the experimental
results of this study.

\vspace{-.5cm}
\bibliographystyle{plain}

\bibliography{biblio}

\begin{thebibliography}{10}

\bibitem{blanquart2016assisted}
Samuel Blanquart, Jean-St{\'e}phane Varr{\'e}, Paul Guertin, Amandine Perrin,
  Anne Bergeron, and Krister~M Swenson.
\newblock Assisted transcriptome reconstruction and splicing orthology.
\newblock {\em BMC Genomics}, 17(10):157, 2016.

\bibitem{brendel2004gene}
Volker Brendel, Liqun Xing, and Wei Zhu.
\newblock Gene structure prediction from consensus spliced alignment of
  multiple ests matching the same genomic locus.
\newblock {\em Bioinformatics}, 20(7):1157--1169, 2004.

\bibitem{burset2000analysis}
M~Burset, IA~Seledtsov, and VV~Solovyev.
\newblock Analysis of canonical and non-canonical splice sites in mammalian
  genomes.
\newblock {\em Nucleic Acids Research}, 28(21):4364--4375, 2000.

\bibitem{christinat2013transcript}
Yann Christinat and Bernard~ME Moret.
\newblock A transcript perspective on evolution.
\newblock {\em IEEE/ACM Transactions on Computational Biology and
  Bioinformatics}, 10(6):1403--1411, 2013.

\bibitem{cunningham2015ensembl}
Fiona Cunningham, M~Ridwan Amode, Daniel Barrell, et~al.
\newblock Ensembl 2015.
\newblock {\em Nucleic Acids Research}, 43(D1):D662--D669, 2015.

\bibitem{edgar2004muscle}
Robert~C Edgar.
\newblock Muscle: multiple sequence alignment with high accuracy and high
  throughput.
\newblock {\em Nucleic Acids Research}, 32(5):1792--1797, 2004.

\bibitem{engstrom2013systematic}
P{\"a}r~G Engstr{\"o}m, Tamara Steijger, Botond Sipos, Gregory~R Grant,
  Andr{\'e} Kahles, Gunnar R{\"a}tsch, Nick Goldman, Tim~J Hubbard, Jennifer
  Harrow, Roderic Guig{\'o}, et~al.
\newblock Systematic evaluation of spliced alignment programs for rna-seq data.
\newblock {\em Nature methods}, 10(12):1185--1191, 2013.

\bibitem{gelfand1996gene}
Mikhail~S Gelfand, Andrey~A Mironov, and Pavel~A Pevzner.
\newblock Gene recognition via spliced sequence alignment.
\newblock {\em Proceedings of the National Academy of Sciences},
  93(17):9061--9066, 1996.

\bibitem{kapustin2008splign}
Yuri Kapustin, Alexander Souvorov, Tatiana Tatusova, and David Lipman.
\newblock Splign: algorithms for computing spliced alignments with
  identification of paralogs.
\newblock {\em Biology direct}, 3(1):20, 2008.

\bibitem{keren2010alternative}
Hadas Keren, Galit Lev-Maor, and Gil Ast.
\newblock Alternative splicing and evolution: diversification, exon definition
  and function.
\newblock {\em Nature Reviews Genetics}, 11(5):345--355, 2010.

\bibitem{CLCview}
B~Knudsen, T~Knudsen, M~Flensborg, H~Sandmann, M~Heltzen, A~Andersen,
  M~Dickenson, J~Bardram, PJ~Steffensen, S~M{\o}nsted, et~al.
\newblock Clc sequence viewer.
\newblock {\em A/S Cb, version}, 6(2), 2011.

\bibitem{kuitche_2016reconstructing}
Esaie Kuitche, Manuel Lafond, and A{\"\i}da Ouangraoua.
\newblock Reconstructing protein and gene phylogenies by extending the
  framework of reconciliation.
\newblock {\em Proceedings of International Conference on Bioinformatics and
  Computational Biology (BICOB'17)}, (ISBN:9781510836679):79--86, 2017.

\bibitem{li2003orthomcl}
Li~Li, Christian~J Stoeckert, and David~S Roos.
\newblock {\em Genome research}, 13(9):2178--2189, 2003.

\bibitem{ranwez2011}
Vincent Ranwez, S{\'e}bastien Harispe, Fr{\'e}d{\'e}ric Delsuc, and Emmanuel~JP
  Douzery.
\newblock {MACSE: Multiple Alignment of Coding SEquences} accounting for
  frameshifts and stop codons.
\newblock {\em PLoS One}, 6(9):e22594, 2011.

\bibitem{sparks2005incorporation}
Michael~E Sparks and Volker Brendel.
\newblock Incorporation of splice site probability models for non-canonical
  introns improves gene structure prediction in plants.
\newblock {\em Bioinformatics}, 21(Suppl\_3):iii20--iii30, 2005.

\bibitem{wu2005gmap}
Thomas~D Wu and Colin~K Watanabe.
\newblock Gmap: a genomic mapping and alignment program for mrna and est
  sequences.
\newblock {\em Bioinformatics}, 21(9):1859--1875, 2005.

\bibitem{zambelli2010}
Federico Zambelli, Giulio Pavesi, Carmela Gissi, David~S Horner, and Graziano
  Pesole.
\newblock Assessment of orthologous splicing isoforms in human and mouse
  orthologous genes.
\newblock {\em BMC Genomics}, 11(1):1, 2010.

\end{thebibliography}

\newpage
\section*{Appendix}

 \begin{figure}[h]
\centering
\includegraphics[height=5.5cm]{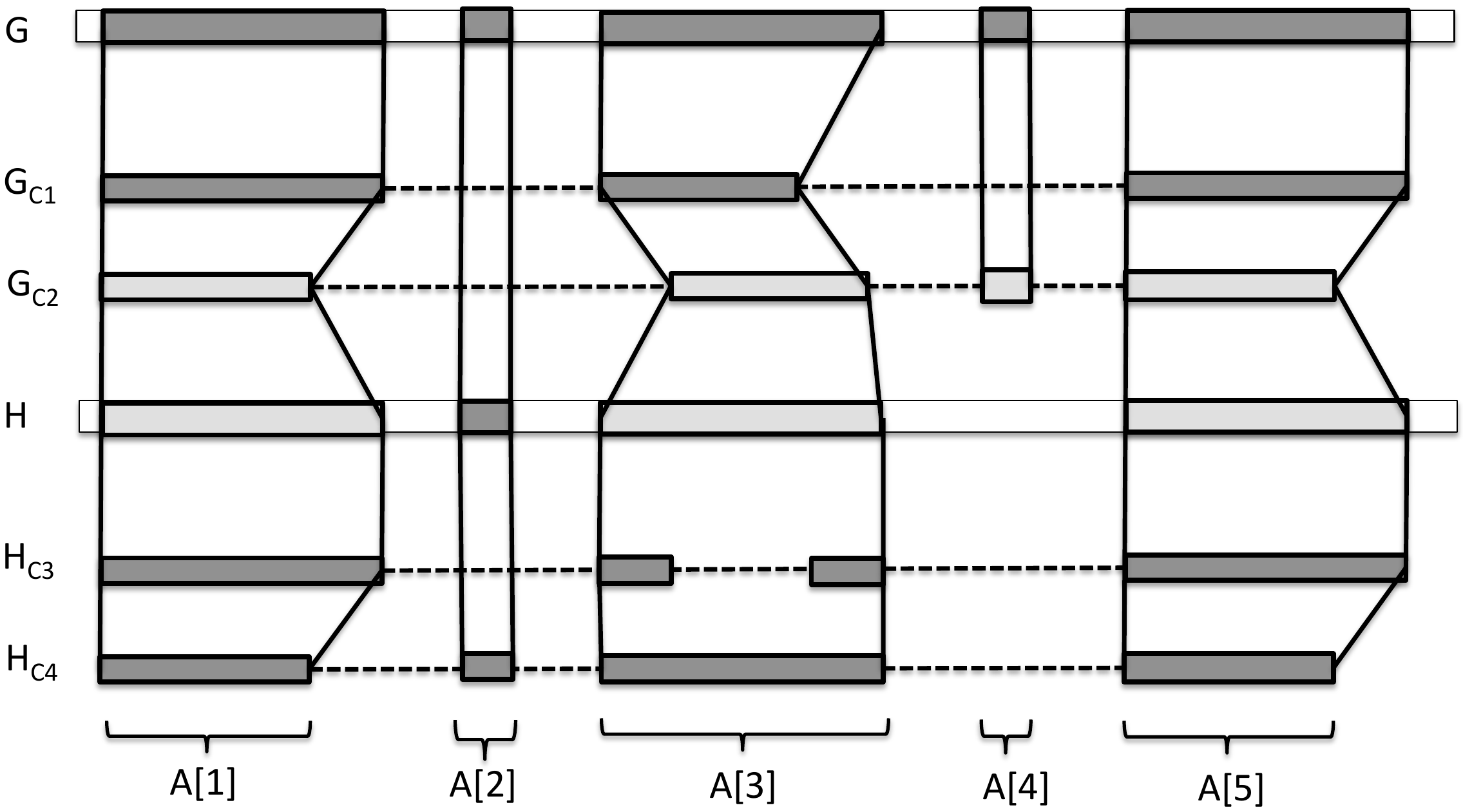}
\caption{Illustration of a multiple spliced alignment of
  $4$ CDS sequences, $G_{C_1},G_{C_2}$ of a gene $G$ and $H_{C_3},H_{C_4}$ of a gene $H$
  on the genes $G$ and $H$. The multiple spliced alignment is composed
  of $5$ blocks such that $A[1],A[3],A[5]$ have occurrences in all sequences,
  while $A[2]$ is conserved in $G$, $H$ and $H_{C_4}$ and $A[4]$ is only
  present in $G$ and $G_{C_2}$. The pairwise spliced alignment induced
  for the CDS sequence $G_{C_2}$ on the gene sequence $H$ is depicted in
  gray color.}
\label{fig:multiplespliced}
\end{figure}

\begin{table}[ht!]

\caption{Detailed description of the two sets of genes and CDS used for the application.}
\begin{tabular}{ll|l|l|l|l|l}
\cline{3-6}
                      &                                       & \multicolumn{4}{c|}{\textbf{Family}}                                                                                                                                                                                                                                                                                                                                                                                                                                                                                                        &  \\ \cline{3-6}
                      &                                       & \multicolumn{2}{c|}{\textbf{FAM86}}                                                                                                                                          & \multicolumn{2}{c|}{\textbf{MAG}}                                                                                                                                             \\ \cline{2-6}
\multicolumn{1}{l|}{} & \multicolumn{1}{c|}{\textbf{Species}} & \multicolumn{1}{c|}{\textbf{Gene\_ID}}                                                      & \multicolumn{1}{c|}{\textbf{\#CDS}}  & \multicolumn{1}{c|}{\textbf{Gene\_ID}}                                                     & \multicolumn{1}{c|}{\textbf{\#CDS}} & \\ \cline{2-6}
\multicolumn{1}{l|}{} & Human                                 & \begin{tabular}[c]{@{}l@{}}ENSG00000158483\\ ENSG00000186523\\ ENSG00000145002\end{tabular} & \begin{tabular}[c]{@{}l@{}}3\\ 4\\ 2\end{tabular}                              & \begin{tabular}[c]{@{}l@{}}ENSG00000105492\\ ENSG00000142512\\ ENSG00000105695\end{tabular} & \begin{tabular}[c]{@{}l@{}}6\\ 7\\ 4\end{tabular}                               &         \\ \cline{2-6}
\multicolumn{1}{l|}{} & Chimpanzee                            & ENSPTRG00000007738                                                                          & 1                                                                              & ENSPTRG00000011374                                                                          & 1                                                                               &                                                                             \\ \cline{2-6}
\multicolumn{1}{l|}{} & Mouse                                 & ENSMUSG00000022544                                                                          & 1                                                                              & ENSMUSG00000051504                                                                          & 4                                                                               &                                                                               \\ \cline{2-6}
\multicolumn{1}{l|}{} & Rat                                   & ENSRNOG00000002876                                                                          & 1                                                                              & ENSRNOG00000021023                                                                          & 2                                                                               &                                                                               \\ \cline{2-6}
\multicolumn{1}{l|}{} & Cow                                   & ENSBTAG00000008222                                                                          & 1                                                                              & ENSBTAG00000017044                                                                          & 1                                                                               &                                                                           \\ \cline{2-6}
\multicolumn{1}{l|}{} & Chiken                                & ENSGALG00000002044                                                                          & 1                                                                              &                                                                                             &                                                                                 &                                                                             \\ \cline{2-6}
\multicolumn{1}{l|}{} & Lizard                                &                                                                                             &                                                                                & ENSACAG00000005408                                                                          & 1                                                                               &                                                                                                \\ \cline{2-6}
\multicolumn{1}{l|}{} & Total                                
& 8 genes                                                                                            & 14                                                                                & 8 genes                                                                         &  26                                                                              &                                                                                                \\ \cline{2-6}
\end{tabular}
\\
For each gene family, the family identifier, the Ensembl identifier of the genes, the number of complete CDS for each gene and the species of genes are given.
\label{familiesdetails}
\end{table}

\begin{table}[ht!]

\caption{Results for the comparison of SFA with the pairwise spliced alignment method SPlign \cite{kapustin2008splign}.}

\begin{tabular}{lllllllllllll}
                      &                                                            &                                                                                              &                                                                                               &                                    &                                                                                             &                                                                                           &                                    &                                                                                           &                                                                                           &                                    &  \\ \cline{3-5}
                      & \multicolumn{1}{l|}{}    & \multicolumn{1}{c|}{\textbf{Method}}                                   & \multicolumn{2}{c|}{\textbf{Family}}                                                                                                                                                                                                                                                  \\ \cline{4-5}
                      & \multicolumn{1}{l|}{} &                                      & \multicolumn{1}{|c|}{\textbf{FAM86}}                                                          & \multicolumn{1}{c|}{\textbf{MAG}} \\ \cline{2-5} 
                      
\multicolumn{1}{l|}{} & \multicolumn{1}{l|}{\textbf{(A)}}         & \multicolumn{1}{c|}{\begin{tabular}[c]{@{}l@{}}\textbf{SFA\_0.0}\\\textbf{SFA\_0.6}\\ \textbf{SFA\_0.7}\\ \textbf{SFA\_0.75}
\\ \textbf{SPlign\_0.0}\\ \textbf{SPlign\_0.75}\end{tabular}}   &

\multicolumn{1}{l|}{\begin{tabular}[c]{@{}l@{}}0.91
\\0.80\\ 0.71\\ 0.65 \\ 0.75 \\  0.40 \end{tabular}}& 

\multicolumn{1}{l|}{\begin{tabular}[c]{@{}l@{}}0.42
\\0.33\\ 0.24\\ 0.22 \\ 0.32 \\  0.20 \end{tabular}} 
\\ \cline{2-5}

\multicolumn{1}{l|}{} & \multicolumn{1}{l|}{\textbf{(B)}}    & \multicolumn{1}{c|}{\begin{tabular}[c]{@{}l@{}}\textbf{SFA\_0.0}\\
\textbf{SFA\_0.6}\\ \textbf{SFA\_0.7}\\ \textbf{SFA\_0.75}
\\ \textbf{SPlign\_0.0}\\ \textbf{SPlign\_0.75}\end{tabular}}  &\multicolumn{1}{l|}{\begin{tabular}[c]{@{}l@{}} 0.56
\\0.55\\ 0.56\\  0.56\\  0.51 \\   0.59 \end{tabular}}
& \multicolumn{1}{l|}{\begin{tabular}[c]{@{}l@{}}0.45
\\ 0.47\\0.53\\ 0.53\\  0.38 \\   0.54 \end{tabular}}
\\ \cline{2-5}

\multicolumn{1}{l|}{} & \multicolumn{1}{l|}{\textbf{(C)}} & \multicolumn{1}{c|}{\begin{tabular}[c]{@{}l@{}}\textbf{SFA\_0.0}\\\textbf{SFA\_0.6}\\ \textbf{SFA\_0.7}\\ \textbf{SFA\_0.75}
\\ \textbf{SPlign\_0.0}\\ \textbf{SPlign\_0.75}\end{tabular}} &

\multicolumn{1}{l|}{\begin{tabular}[c]{@{}l@{}}0.79
\\0.77\\  0.76\\  0.75 \\ 0.71 \\   0.74 \end{tabular}}
& \multicolumn{1}{l|}{\begin{tabular}[c]{@{}l@{}}0.77
\\ 0.80\\  0.88\\ 0.90 \\ 0.62 \\  0.89 \end{tabular}}
\\ \cline{2-5}

\multicolumn{1}{l|}{} & \multicolumn{1}{l|}{\textbf{(D)}}         & \multicolumn{1}{c|}{\begin{tabular}[c]{@{}l@{}}\textbf{SFA\_0.0}\\\textbf{SFA\_0.6}\\ \textbf{SFA\_0.7}\\ \textbf{SFA\_0.75}
\\ \textbf{SPlign\_0.0}\\ \textbf{SPlign\_0.75}\end{tabular}}    &

\multicolumn{1}{l|}{\begin{tabular}[c]{@{}l@{}} 0.94
\\0.93\\ 0.93\\  0.92 \\  0.88 \\ 0.96 \end{tabular}}
& \multicolumn{1}{l|}{\begin{tabular}[c]{@{}l@{}}0.90
\\  0.91\\  0.95 \\ 0.96 \\  0.75 \\   0.98 \end{tabular}}
\\ \cline{2-5}
\end{tabular}
\label{tab:structure}
\\ For each method and value of the min\_idty parameter, the number correspond to
(A) the overall CDS coverage of the spliced alignments and the ratio
of internal block extremities (all block extremities except the start 
of the first block and the end
of the last block in a spliced alignment) that correspond
to (B) an actual CDS exon extremity, (C) an 
actual gene exon extremity and  (D) a canonical splice site are given.
\end{table}

\begin{sidewaysfigure}
\vspace{-11cm}
 \includegraphics[width=\textheight]{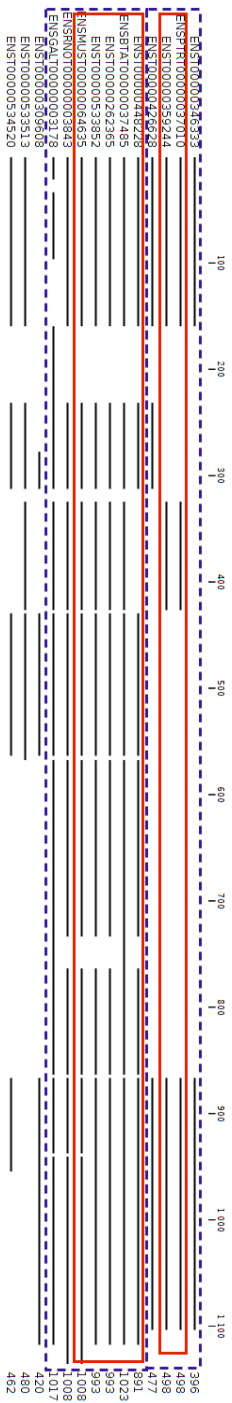}\vspace{-11cm}
\caption{The sets of ortholog clusters obtained using the SFA and the OrthoMCL methods for the FAM86 family. The clusters containing more 
than one CDS are depicted in red color for the SFA method, 
and in dotted blue colored line for OrthoMCL. The visualization 
of the alignment was generated using CLC Sequence Viewer \cite{CLCview}}.
\end{sidewaysfigure}
\newpage
\begin{sidewaysfigure}
\label{fig:fam86_clusters}
\vspace{-11cm}
 \includegraphics[width=\textheight]{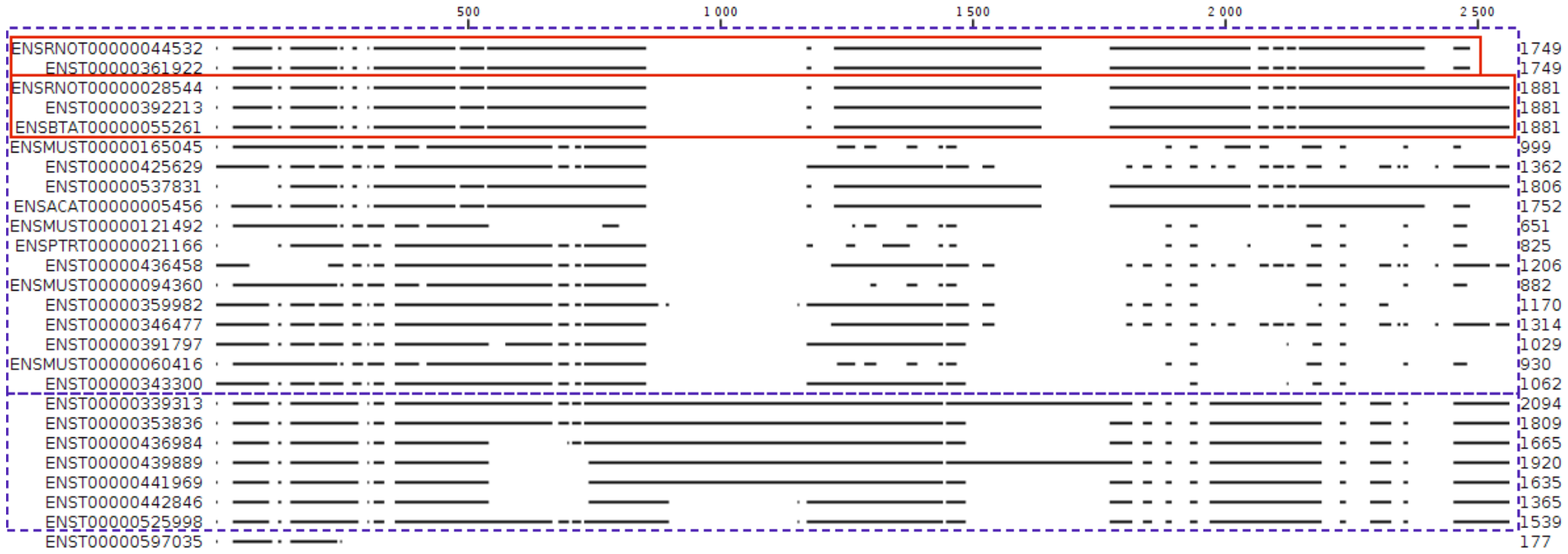}
\vspace{-9cm}
    \caption{The sets of ortholog clusters obtained using the SFA and the OrthoMCL methods for the MAG family.}
  \label{fig:mag_clusters}
\end{sidewaysfigure}

\begin{table}[ht!]
\caption{Results for the comparison of SFA with the method MACSE \cite{ranwez2011} for multiple CDS alignment.}
\label{msaevaluation}
\begin{tabular}{llll}
\hline
\multicolumn{ 1}{|l|}{\textbf{Gene family}}            & \multicolumn{1}{l|}{\textbf{Methods}} & \multicolumn{1}{l|}{\textbf{(A)}} & \multicolumn{1}{c|}{\textbf{(B)}} \\ \hline \hline

\multicolumn{1}{|l|}{\multirow{6}{*}{FAM86}} 
& \multicolumn{1}{l|}{\textbf{SFA\_0.0}}   & \multicolumn{1}{l|}{0.49}                    & \multicolumn{1}{l|}{ 0.81}   \\ \cline{2-4}

\multicolumn{1}{|l|}{}                       & \multicolumn{1}{l|}{\textbf{SFA\_0.5}}   & \multicolumn{1}{l|}{0.49}                    & \multicolumn{1}{l|}{0.81}   \\ \cline{2-4}

\multicolumn{1}{|l|}{}                       & \multicolumn{1}{l|}{\textbf{SFA\_0.6}}   & \multicolumn{1}{l|}{0.56}                    & \multicolumn{1}{l|}{ 0.86}   \\ \cline{2-4}

\multicolumn{1}{|l|}{}                       & \multicolumn{1}{l|}{\textbf{SFA\_0.7}}   & \multicolumn{1}{l|}{0.82}                    & \multicolumn{1}{l|}{ 0.82}   \\ \cline{2-4}

\multicolumn{1}{|l|}{}                       & \multicolumn{1}{l|}{\textbf{SFA\_0.75}}   & \multicolumn{1}{l|}{0.84}                    & \multicolumn{1}{l|}{0.82}   \\ \cline{2-4}

\multicolumn{1}{|l|}{}                       & \multicolumn{1}{l|}{\textbf{MACSE}}   & \multicolumn{1}{l|}{0.47}                    & \multicolumn{1}{l|}{ 0.29}   \\ \hline \hline 

\multicolumn{1}{|l|}{\multirow{6}{*}{MAG}}

& \multicolumn{1}{l|}{\textbf{SFA\_0.0}}   & \multicolumn{1}{l|}{0.63}                    & \multicolumn{1}{l|}{ 0.57}   \\ \cline{2-4}

\multicolumn{1}{|l|}{}                       & \multicolumn{1}{l|}{\textbf{SFA\_0.5}}   & \multicolumn{1}{l|}{0.73}                    & \multicolumn{1}{l|}{ 0.48}   \\ \cline{2-4}

\multicolumn{1}{|l|}{}                       & \multicolumn{1}{l|}{\textbf{SFA\_0.6}}   & \multicolumn{1}{l|}{0.74}                    & \multicolumn{1}{l|}{ 0.65}   \\ \cline{2-4}

\multicolumn{1}{|l|}{}                       & \multicolumn{1}{l|}{\textbf{SFA\_0.7}}   & \multicolumn{1}{l|}{0.92}                    & \multicolumn{1}{l|}{0.84}   \\ \cline{2-4}

\multicolumn{1}{|l|}{}                       & \multicolumn{1}{l|}{\textbf{SFA\_0.75}}   & \multicolumn{1}{l|}{0.93}                    & \multicolumn{1}{l|}{0.80}   \\ \cline{2-4}

\multicolumn{1}{|l|}{}                       & \multicolumn{1}{l|}{\textbf{MACSE}}   & \multicolumn{1}{l|}{0.44}                    & \multicolumn{1}{l|}{0.05}   \\ \hline 
\end{tabular}
\\
For each method, the number
correspond to (A) the percentage of columns identity  in the multiple CDS alignments
and (B) the overall number of long gaps corresponding to real introns divided (\slash)
by the total number of long gaps.
\end{table}

\begin{sidewaysfigure}
\vspace{-11cm}
 \includegraphics[width=\textheight]{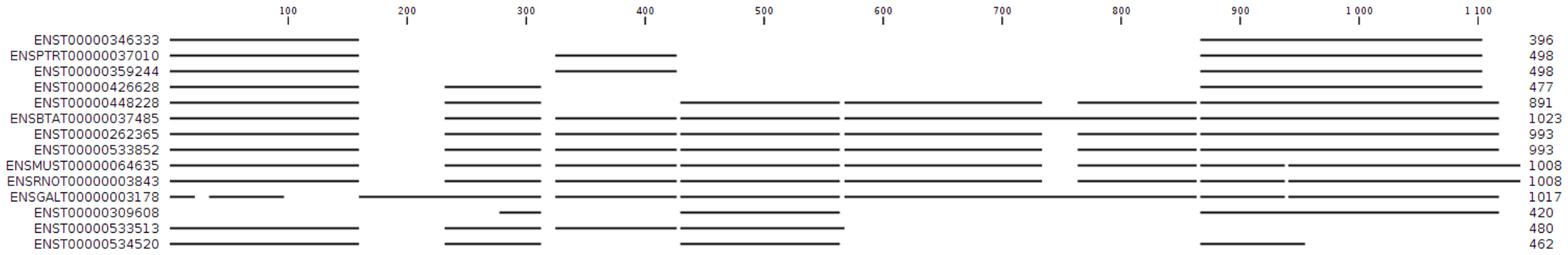}
\vspace{-11cm}
\caption{The multiple CDS alignment obtained using the SFA method for the FAM86 family. The visualization 
of the alignment was generated using CLC Sequence Viewer \cite{CLCview}.}
  \label{fig:fam86_sfa}
\vspace{-9cm}
\includegraphics[width=\textheight]{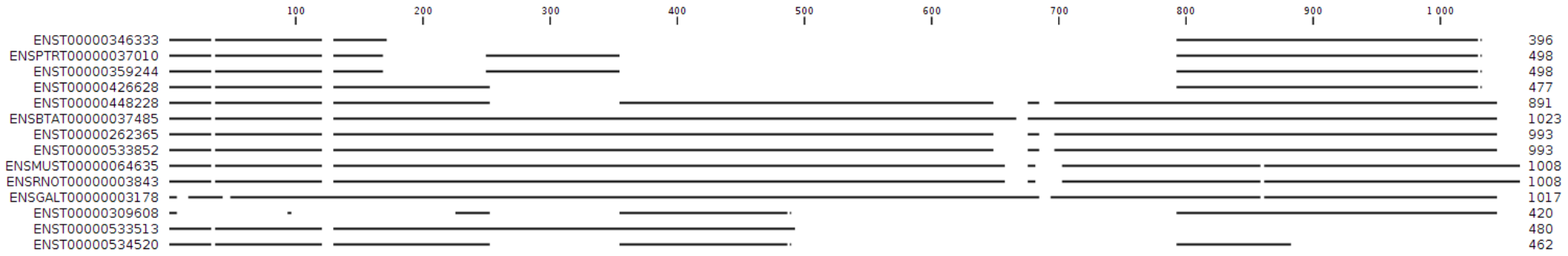}
\vspace{-11cm}
    \caption{The multiple CDS alignment obtained using the MACSE method \cite{ranwez2011} for the FAM86 family.}
  \label{fig:fam86_macse}
\end{sidewaysfigure}


\end{document}